\documentclass[aps,prl,showpacs,amsmath,twocolumn,amssymb,superscriptaddress,letterpaper]{revtex4}
\usepackage{times}
\usepackage{amsfonts}
\usepackage{mathrsfs}
\usepackage{graphicx}
\usepackage{dcolumn}
\usepackage{bm}
\usepackage{color}

\usepackage[colorlinks,bookmarks=false,citecolor=blue,linkcolor=red,urlcolor=blue]{hyperref}
\usepackage{bibunits}

\def\be{\begin{equation}}       \def\ee{\end{equation}}
\def\bea{\begin{eqnarray}}      \def\eea{\end{eqnarray}}

\begin{document}

\begin{bibunit}

\title{ Nematic orders and nematicity-driven topological phase transition in FeSe }

\author{Xianxin Wu}\thanks{These authors contributed equally to this work.}
\affiliation{ Institute of Physics, Chinese Academy of Sciences,
Beijing 100190, China}

\author{Yi Liang}\thanks{These authors contributed equally to this work.}
\affiliation{ Graduate School of Chongqing Normal University,
Chongqing 401331, China}
\affiliation{ Institute of Physics, Chinese Academy of Sciences,
Beijing 100190, China}

\author{Heng Fan }  \affiliation{ Institute of Physics, Chinese Academy of Sciences,
Beijing 100190, China}
\affiliation{Collaborative Innovation Center of Quantum Matter, Beijing, China}

\author{Jiangping Hu  }\email{jphu@iphy.ac.cn} \affiliation{ Institute of Physics, Chinese Academy of Sciences,
Beijing 100190, China}\affiliation{Department of Physics, Purdue University, West Lafayette, Indiana 47907, USA}
\affiliation{Collaborative Innovation Center of Quantum Matter, Beijing, China}


\date{\today}

\begin{abstract}
 We investigate  nematic states in both bulk FeSe and FeSe thin films. It is found that their band structures and  signature features that were observed in a variety of experiments can be perfectly explained by introducing the $d$-wave nematic orders that are required to have  contributions from all $t_{2g}$ d-orbitals, which contradicts the conventional wisdom that the nematicity is simply driven by the orbital degeneracy between the $d_{xz}$ and $d_{yz}$ orbitals. These orders can be generated by the Coulomb interaction between the nearest neighbor Fe sites.  In the presence of spin-orbital couplings,  we predict that the nematic order can drive a topological quantum phase transition through a band inversion at  the $M$ point of the Brillouin zone   to  produce topologically protected edge states near the Fermi level. The prediction makes FeSe as a tunable system to integrate topological properties into high temperature superconductivity to realize Majorana related physics.
\end{abstract}


\maketitle

The nematicity, which breaks rotational symmetries but preserves translational symmetry of lattice, is one of the most intriguing properties in iron based superconductors\cite{Fang2008,Xu2008,Fernandes2014,Chu2012}.  The microscopic origin of the nematicity  in these materials has been debated intensively as the evidences supporting both magnetic and orbital based mechanisms exist.  Understanding the nematic origin  can also help to understand  superconducting states\cite{Fernandes2014}.

The nematicity in the structurally simplest iron-based superconductor FeSe is particularly interesting. The bulk FeSe undergoes a tetragonal-to-orthorhombic structural phase transition at $T_s\sim90$ K and exhibits superconductivity at 8 K. A significant enhancement of T$_c$ can be achieved  under external pressure\cite{Margadonna2009} or on monolayer FeSe grown on SrTiO$_3$ surfaces\cite{Wang2012,LiuDF2012,He2012,Tan2012}. The nematic order  coexists with superconductivity but not with long-range magnetic order.  The absence of the long-range magnetic order has led to arguments that  the origin of the nematicity is not magnetically driven and is most likely orbital-driven. For example, experimentally, in NMR measurements, spin-lattice relaxation rate is found to be not affected at nematic temperature, which favors an orbitally driven nematic behavior in FeSe\cite{Baek2015,Bohmer2015}.

 However, recent theoretical calculations and experiments show that  the nematic states  in the FeSe systems are far more intriguing and complex\cite{Chubukov2015,Glasbrenner2015,WangF2015,YuR2015,QSWang2015}. There are very strong high energy spin fluctuations\cite{QSWang2015} which suggest that the nematicity and magnetism may be still intimately linked. It was also found that there are many interesting features in the band structures of the nematic state.  In both bulk FeSe and FeSe thin films, the band splitting between $d_{xz}$ and $d_{yz}$ bands at $\Gamma$ is temperature-insensitive but the  splitting at $M$ is closely related to the structural phase transition and can reach 80 meV\cite{Zhang2015,ZhangY2015}. The former could be attributed to the spin-orbit coupling and the latter can be  attributed to  the nematicity\cite{Zhang2015,Watson2015,Mukherjee2015}. The Dirac cone type of band dispersions around $M$ point was observed in FeSe thin films thicker than 1 Unit Cell \cite{Tan2015,Li2015}. The corresponding Fermi surfaces around $M$ are four propeller-like electron pockets. These features  can not be fully explained by  an onsite ferro-orbital  and a $d$-wave orbital ordering within $d_{xz/yz}$ orbitals\cite{Zhang2015}. Thus, it is of great importance to figure out  orders in nematic phase that can consistently explain all these features. Such an understanding can shed light on the origin of nematicity and the mechanism of superconductivity in FeSe.

Another interesting issue in FeSe systems is the   existence of possible nontrivial topology\cite{Hao2014,Wu2014,Wang2015} .  Theoretically, nontrivial topology has been  predicted  in the Fe(Te,Se) systems in a variety of circumstances, which suggests that FeSe can be intriguing systems to integrate topological physics together with high T$_c$ superconductivity. However, the topological properties in these  predictions are untunable for  given materials and it is also unknown that how  the nematicity can affect the topological properties.

In this paper, we investigate the nematic order in FeSe systems including both bulk FeSe and FeSe thin films. We find that  the band structures in ARPES experiments can be perfectly understood if the $d$-wave orbital orders emerge in all $t_{2g}$ orbitals, namely $d_{xz/yz}$ and $d_{xy}$ orbitals.  The order from $d_{xy}$ orbital  is even strongest. This result is sharply against the conventional wisdom that the nematicity is driven by the degeneracy between $d_{xz}$ and $d_{yz}$ orbitals. We show that  the Coulomb interactions between nearest neighbor sites  can produce these $d$-wave orders. The Dirac cones around M point mentioned above are attributed to all $d_{xz/yz}$ and $d_{xy}$ orbitals and they exist in the normal state but are pushed up to near the Fermi level in the nematic phase. Furthermore, we predict that the nematicity can drive a topological phase transition through a band inversion at $M$ point. The strength of the nematic order can be considered as an external tunable parameter to control topological properties, which makes FeSe  a tunable system to integrate topological properties into high temperature superconductivity to realize Majorana related physics.

\begin{figure}[t]
\centerline{\includegraphics[height=5cm]{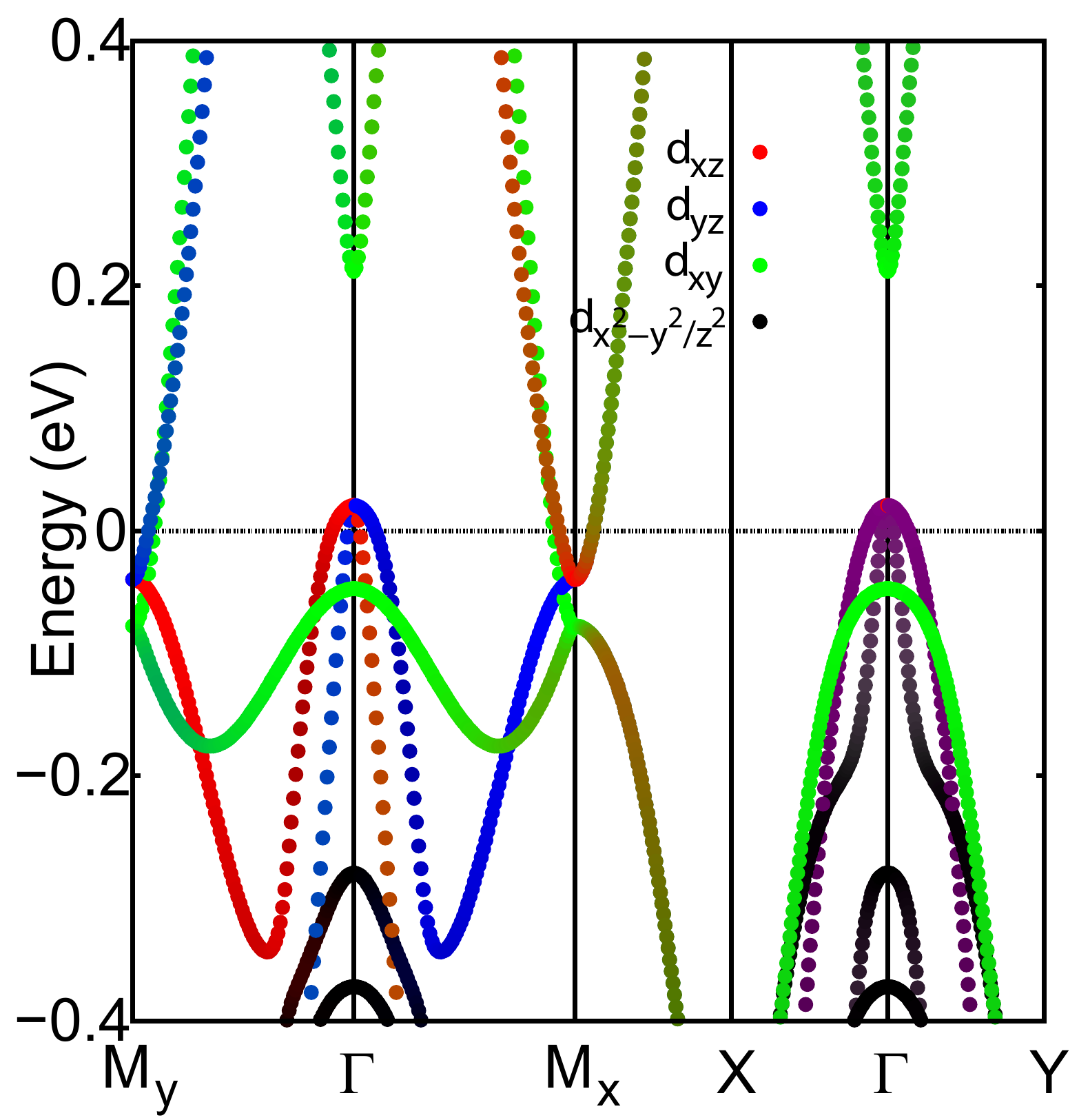}}
\caption{(color online). The band structure of FeSe with renormalization and additional shifts to hopping parameters. The oribital characters are indicated by different colors. \label{band} }
\end{figure}

{\em Tight binding model}
To investigate the nematic order, we start from a tight binding model for the FeSe systems. For monolayer FeSe, the band structure can be fully unfolded into the Brillouin Zone of an Fe square lattice. The five-band tight binding model in momentum space with respect to one Fe unit cell can be written as,
\begin{eqnarray}
H_t=\sum_{{\sigma},k\in BZ1}\phi^{\dag}_{\sigma}(k)A(k)\phi_{\sigma}(k).
\end{eqnarray}
where $\psi^{\dag}_{\sigma}(\mathbf{k}) =[c_{\mathbf{k}1\sigma}^{\dag},c_{\mathbf{k}2\sigma}^{\dag},c_{\mathbf{k+Q}3\sigma}^{\dag},c_{\mathbf{k+Q}4\sigma}^{\dag},c_{\mathbf{k+Q}5\sigma}^{\dag}]$
, $\mathbf{Q}=(\pi,\pi)$ and BZ1 denotes the Brillouin zone of one Fe lattice. Here, the Fe $3d$ orbitals are denoted by numbers, i.e., $(1,2,3,4,5)\rightarrow(xz,yz,x^2-y^2,xy,z^2)$. The matrix elements of $A(k)$ can be found in the supplementary material. It is well known that there are some differences between the bands from DFT and those in ARPES experiment for FeSe. To best fit the bands in experiment, we apply the renormalization from interatomic Coulomb interaction and additional shifts to hopping parameters. The final band shown in Fig.\ref{band} can be classified into ``$k$" band and ``$k+Q$" band\cite{Wu2014} and the corresponding hopping parameters are given in supplementary material. Above T$_s$, the $d_{xz/yz}$ bands are slightly above $E_F$ and the flat $d_{xy}$ band locates at -50 meV at $\Gamma$ point. The $d_{xz/yz}$ bands are present at -40 meV and the $d_{xy}$ bands locate at -80 meV at $M$ point. All these features can be quantitatively compared with the bands  measured in experiment for mulitlayer or bulk FeSe\cite{Tan2013,Watson2015,Zhang2015}.

\begin{figure}[t]
\centerline{\includegraphics[height=4 cm]{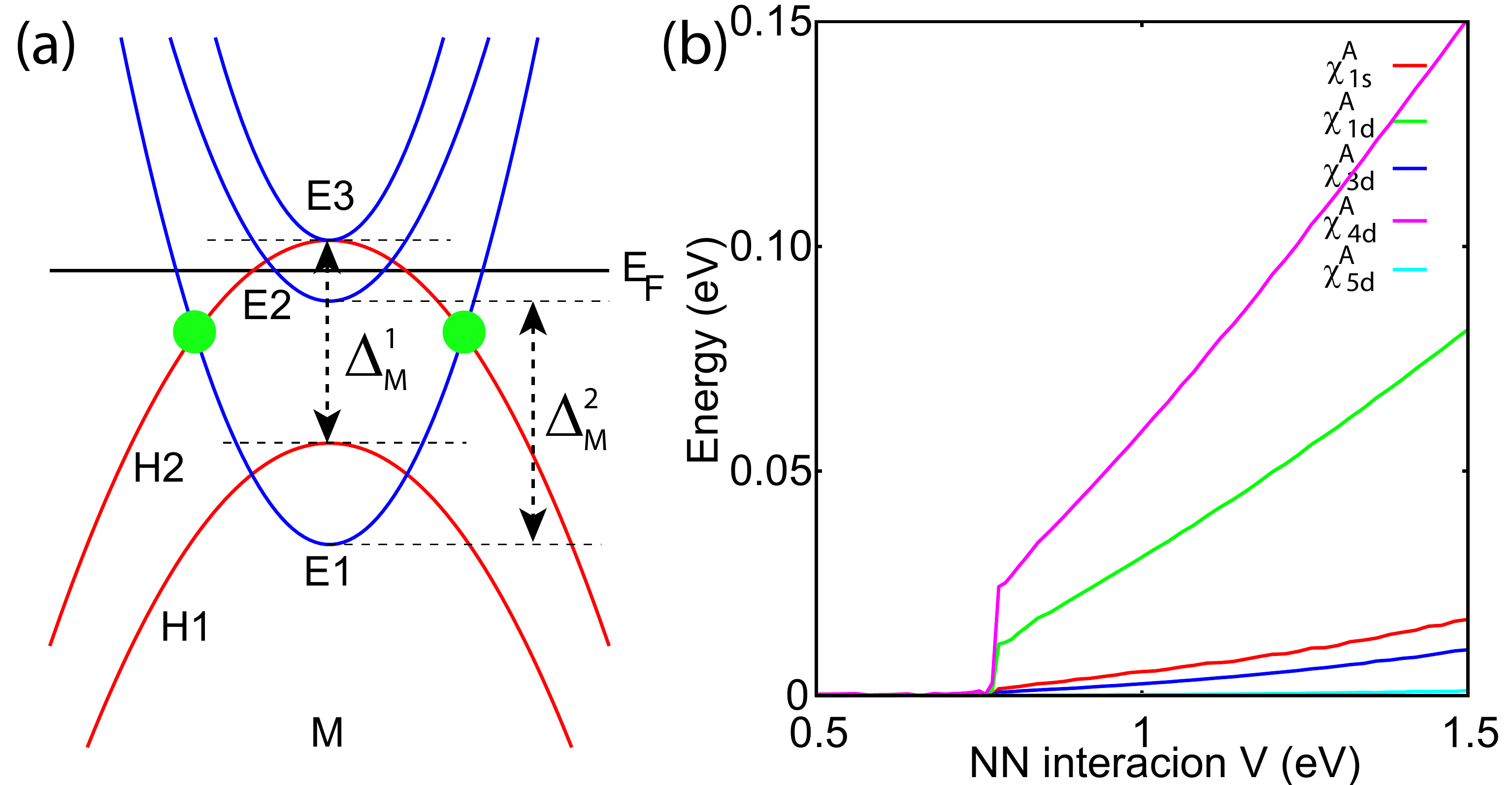}}
\caption{(color online). (a) Band structure for mutilayer and bulk FeSe in nematic phase in ARPES experiments. The green circles represent the Dirac points. (b) Symmetry-breaking order parameters as a function of interatomic Coulomb interaction $V$.  \label{bandexp} }
\end{figure}

{\em Band structure in nematic phase} We first summarize the important features in the band structure of the nematic states observed in  many ARPES experiments  on  the multilayer or  bulk FeSe. The most striking feature at $\Gamma$ point is the splitting between $d_{xz}$ and $d_{yz}$ bands, which is nearly temperature independent and persists above $T_{nem}$\cite{Zhang2015,ZhangY2015}. It goes against the ferro-orbital ordering and may be explained by spin-orbit coupling in FeSe. Furthermore, the band splitting at $M$ point deceases with the increasing of temperature and finally vanishes at a certain temperature above $T_S$. The bands at zone corner are much more complicated compared with those at $\Gamma$ point. As the band splitting at $M$ is attributed to the nematicity, we focus on the bands around $M$. Two hole-like bands at $M$, labelled as H1 and H2  respectively   in red colors  shown in Fig.\ref{bandexp}(a),  have been observed in all experiments. A deep electron band crossing with the bottom H1 hole band have been identified in both multilayer and bulk FeSe\cite{Watson2015,Tan2013}. A shallow electron band, which is very close to the Fermi level (FL) and crosses with the top H2 hole band, has also been observed\cite{Zhang2015,Tan2013}. At low temperature, an electron band slightly above the FL was observed and its band bottom is degenerate with the band top of the H2 hole band\cite{Zhang2015,Tan2013}. After summarizing all the experimental facts, we conclude that there are two hole bands and three electron bands around M point in ARPES experiments in nematic phase, which is shown in Fig.\ref{bandexp}(a). Moreover, the deep electron band intersects with the top hole band near the FL, forming a Dirac cone\cite{Tan2015,Li2015}(the green circles in Fig.\ref{bandexp}(a)). As the temperature increases, the gap between shallow the E2 electron band (shown in Fig.\ref{bandexp}(a)) and the bottom H1 hole band decreases. A linear dispersion is observed when the two bands finally touch each other\cite{Tan2013}.

\begin{figure}[t]
\centerline{\includegraphics[height=6 cm]{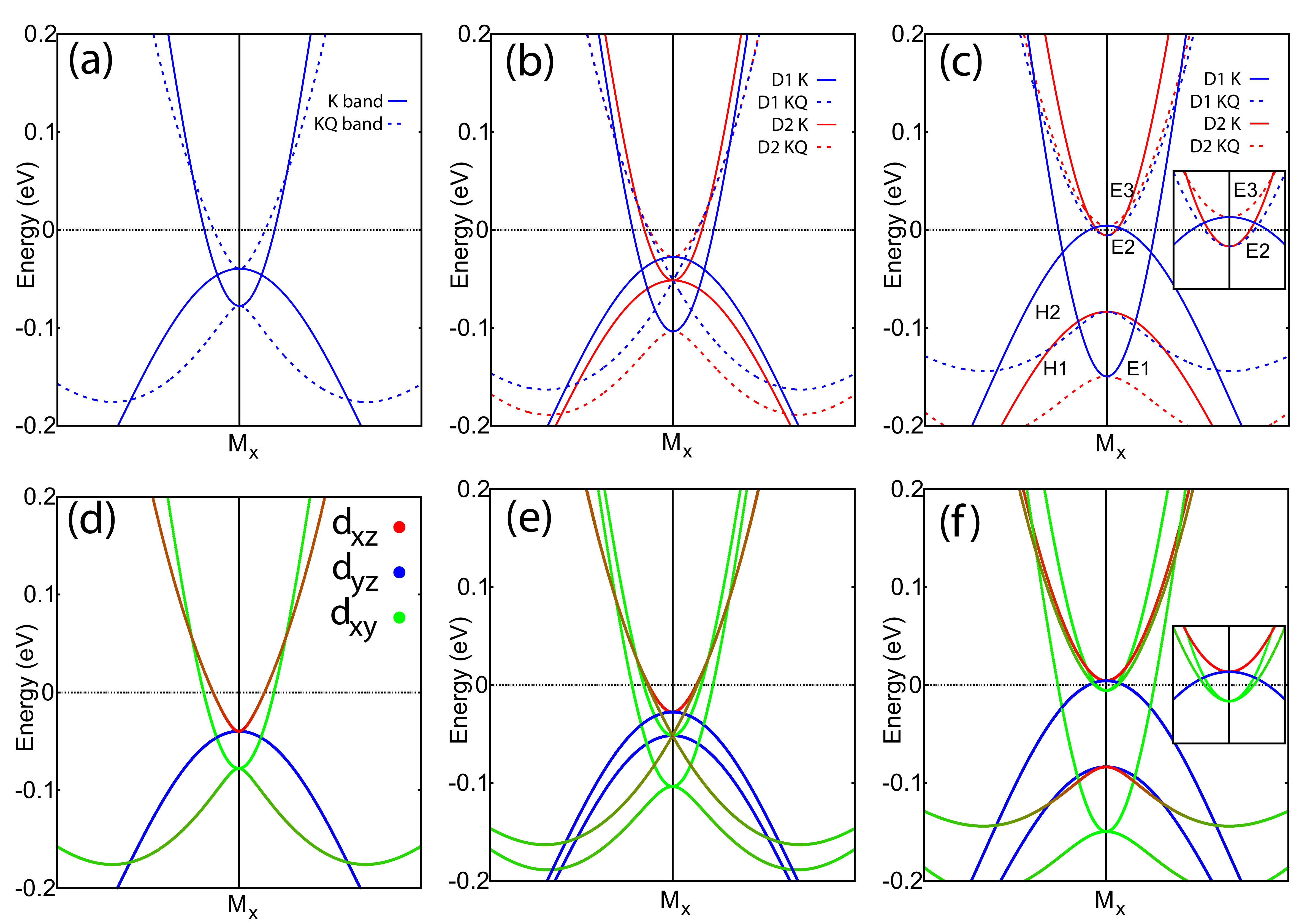}}
\caption{ (color online). The evolution of band structure along $\Gamma-M_x$ direction in two domains with nematic orders. (a),(d) nematic orders vanish. (b),(e) $\chi^A_{1d}=3 $ meV and $\chi^A_{4d}=6.5$ meV. (c),(f) $\chi^A_{1d}=11 $ meV and $\chi^A_{4d}=18$ meV.  The solid lines represent ``$k$" bands and the dashed lines represent ``$k+Q$" bands. The blue and red lines represent the bands of domain1 and domian2, respectively. The insets show the zoom-in bands around M point near the FL. The bands shown in (c) are in good agreement with experiment.
 \label{nematic} }
\end{figure}

 An onsite orbital order for $d_{xz/yz}$ orbitals  is clearly insufficient to reproduce the observed results. Here, we will show that the nematic orders that can be spontaneously generated by the nearest neighbor (NN) interatomic Coulomb interaction for five $d$ orbitals\cite{Hu2015,Jiang2015} can be sufficient to explain experimental results.  The NN interaction can be written as
\begin{eqnarray}
H_V=V\sum_{\langle ij \rangle} :n_{i}n_{j}:,
\end{eqnarray}
where the "normal-order" sign represents that the direct Hartree term depending on the total density $n_i=\sum_\alpha n_{i\alpha}$ is removed. This interaction can be decoupled into two kinds of terms : (1) symmetry-preserving terms, which are corrections to the DFT-based bands; (2) spontaneous symmetry-breaking terms, which are nematic orders from this interaction.  We focus on the symmetry-breaking terms as the symmetry-preserving terms are generally absorbed to make the band structures consistent with experiments. As only $C_4$ rotational symmetry is broken in the nematic phase, we only consider the terms that break $C_4$ symmetry but preserve other symmetries, such as glide-plane symmetry and $C_2$ rotational symmetry around the nearest Fe-Fe bond. The symmetry-breaking Hamiltonian in the mean-field level reads,
\begin{eqnarray}
H^B_{V}&=&2\sum_{k\sigma}[\chi^A_{1s}\alpha_k(c^{\dag}_{k1\sigma}c_{k1\sigma}-c^{\dag}_{k2\sigma}c_{k2\sigma})\nonumber\\
&+&\chi^A_{1d}\beta_k(c^{\dag}_{k1\sigma}c_{k1\sigma}+c^{\dag}_{k2\sigma}c_{k2\sigma})+\chi^A_{3d}\beta_k c^{\dag}_{k3\sigma}c_{k3\sigma}\nonumber\\
&+&\chi^A_{4d}\beta_k c^{\dag}_{k4\sigma}c_{k4\sigma}+\chi^A_{5d}\beta_k c^{\dag}_{k5\sigma}c_{k5\sigma}\nonumber\\
&+&(\chi^A_{13p}\eta^y_kc^{\dag}_{k1\sigma}c_{k+Q3\sigma}+\chi^A_{13p}\eta^x_kc^{\dag}_{k2\sigma}c_{k+Q3\sigma})\nonumber\\
&+&(\chi^A_{14p}\eta^x_k c^{\dag}_{k1\sigma}c_{k+Q4\sigma}-\chi^A_{14p}\eta^y_k c^{\dag}_{k2\sigma}c_{k+Q4\sigma})\nonumber\\
&+&(\chi^A_{15p}\eta^y_k c^{\dag}_{k1\sigma}c_{k+Q5\sigma}- \chi^A_{15p}\eta^x_k c^{\dag}_{k2\sigma}c_{k+Q5\sigma})\nonumber\\
&+&\chi^A_{35s}\beta_k(c^{\dag}_{k3\sigma}c_{k5\sigma}+c^{\dag}_{k5\sigma}c_{k3\sigma})],
\end{eqnarray}
where $\alpha_k=cosk_x+cosk_y$, $\beta_k=cosk_x-cosk_y$ and $\eta^{x/y}_k=isink_{x/y}$. The order parameters are given the supplementary material.

We can  get a self-consistent solution of the Hamiltonian and obtain the symmetry-breaking order parameters as a function of interatomic Coulomb interaction $V$, shown in Fig.\ref{bandexp}(b). We find that the intra-orbital nematic orders in the $d$-wave channels for $d_{xz/yz}$ and $d_{xy}$ orbitals establish from $V\sim0.75$ eV. They are dominant over other nematic orders which are very small with little effect on band structures. Their dominance can be attributed to the fact that  the Fermi level is close to the van Hove singularity (vHS) contributed by $d_{xz/yz}$ and $d_{xy}$ orbitals that can generate large quantum fluctuations\cite{Jiang2015} (see the supplementary material). We also notice in our calculations that the nematic order for $d_{xy}$ is  even slightly stronger than those of $d_{xz/yz}$ orbitals.  This result  is in sharp contrast to previous studies in nematic phase, which suggest that the nematicity is driven by the degeneracy between the $d_{xz}$ and $d_{yz}$ orbitals.

Now we consider the consequences of  the $d$-wave nematic orders. The $d$-wave form factors in the nematic state have several interesting consequences. First, they   vanish at $\Gamma$ point but reach the maximum at $M$ point.  Second, their signs are opposite between $k$ and $k+Q$  points. Namely, the nematic orders have opposite signs between the ``$k$"  and ``$k+Q$" bands. Finally,  in the band structure,  as the  couplings between  $d_{xz/yz}$ and $d_{xy}$ orbitals are specified between momentum $k$  and $k+Q$, the nematic orders  at a specific $k$ point are also effectively opposite between  $d_{xz/yz}$ and $d_{xy}$ orbitals.

These features create specific band structure reorganizations at $M$ point in the nematic state as shown in Fig.\ref{nematic} in the original 2-Fe unit cell. Fig.\ref{nematic}(a) and (d) show the bands and orbital characters along $\Gamma-M$ direction without nematic orders. At M point, the ``$k+Q$" $d_{xz}$ electron band and ``$k$" $d_{yz}$ hole band are degenerate, so do the ``$k+Q$" $d_{xy}$ hole band  and the ``$k$"  $d_{xy}$ electron band. If we consider an undetwinned  nematic state with two nematic domains  characterized by opposite nematic order parameters, in domain1 (D1) with negative nematic order parameters, the ``$k+Q$" $d_{xz}$ electron band shifts down but the ``$k+Q$" $d_{xy}$ hole band shifts up. Such movements  make these two bands  closer at M points as shown in Fig.\ref{nematic}. Simultaneously, the two "$k$" bands move away from each other.  However, opposite behaviors appear in the domain2 (D2): ``$k+Q$" bands move away from each other and the "$k$"  bands move closer. Fig.\ref{nematic}(b) and (e) show the band structure of two domains for the case of $\chi^A_{1d}=3 $ meV and $\chi^A_{4d}=6.5$ meV. We find that the "$k+Q$" bands in D1 touch each other with linear dispersion but the "k" bands in D2 touch quadratically. The former is attributed to the linear coupling between $d_{xz}$ and $d_{xy}$ orbitals and the latter is attributed to the vanish of coupling between $d_{yz}$ and $d_{xy}$ orbitals along $\Gamma-M_x$ direction. With the nematic order further increasing, in D1 the two "$k$" bands move away and the two "$k+Q$" bands couple with each other away from $M$ point, opening a gap. With $\chi^A_{1d}=11 $ meV and $\chi^A_{4d}=18$ meV, the band structure of two domains are shown in Fig.\ref{nematic}(c) and (f). The band in Fig.\ref{nematic}(c) agrees extremely well with that in ARPES experiment shown in Fig.\ref{bandexp}(a). Furthermore, the E2 and H1 bands touch each other with linear dispersion as nematic orders weaken, which is also consistent with experimental observations\cite{Tan2013}. The two bands touching quadratically in D2 has been not observed, which is likely attributed to the  matrix element effect in APRES measurements. We expect them to be observed around $M_y$ point as they belong to the "$k$" bands.

\begin{figure}[t]
\centerline{\includegraphics[height=4 cm]{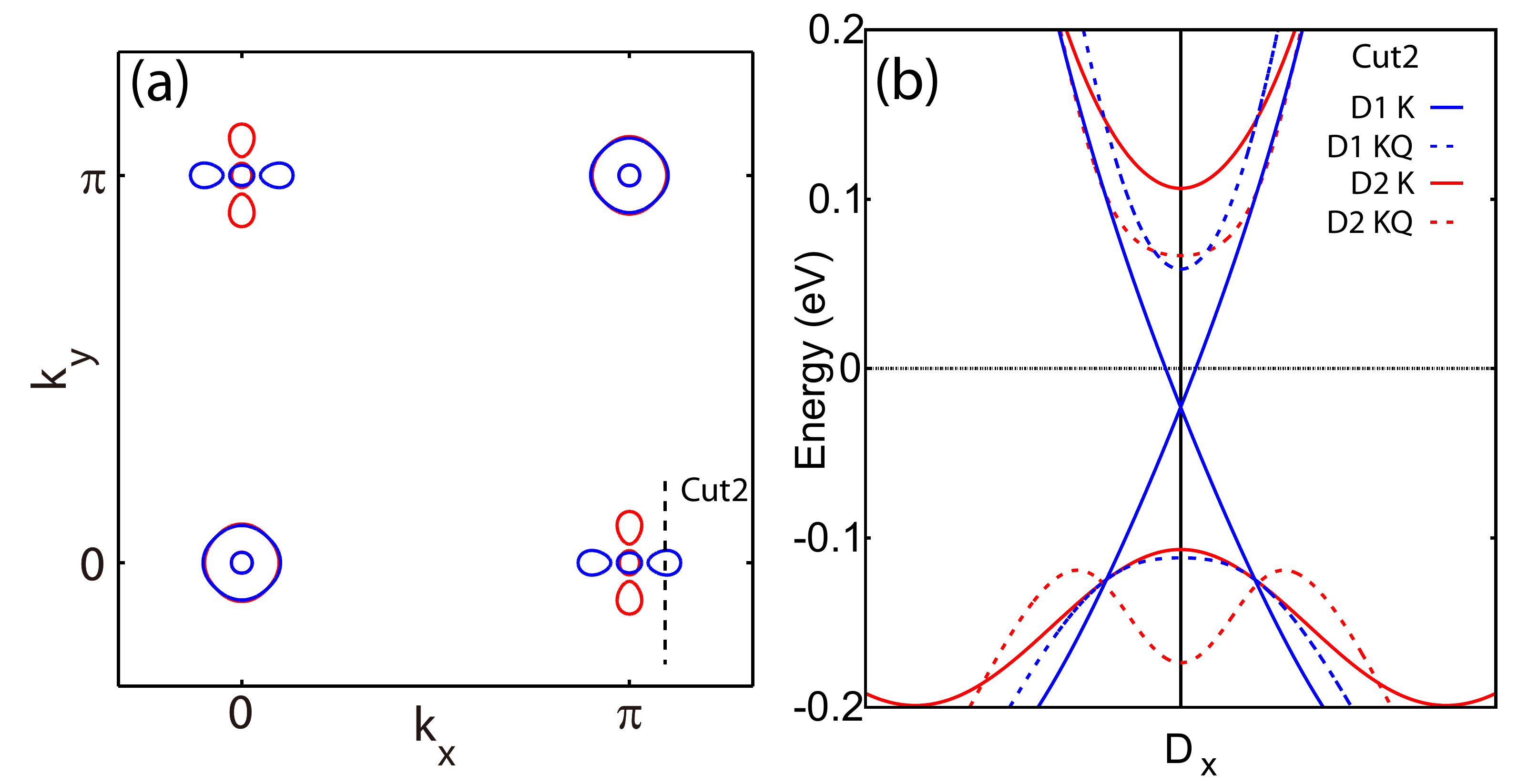}}
\caption{ (color online). (a) Fermi surfaces for two domains with $\chi^A_{1d}=11 $ meV and $\chi^A_{4d}=18$ meV. (b) Band structure along the $k_y$ cut (Cut2 in (a)) across the Dirac cone near M point. The blue and red lines represent the bands of domain1 and domian2, respectively.
 \label{FSdirac} }
\end{figure}

The Fermi surfaces with $\chi^A_{1d}=11 $ meV and $\chi^A_{4d}=18$ meV, corresponding to the band in Fig.\ref{nematic}(c), are given in Fig.\ref{FSdirac}(a). In the figure, the four propeller-like electron pockets around $M_x$ point are attributed to the Dirac cones, which are contributed from the $d_{xz/yz}$ and $d_{xy}$ orbitals. The small hole pockets centering $M$ is attributed to $d_{xy}$ orbitals. All these Fermi surfaces are fully consistent with the experimental observations in the nematic phase\cite{ZhangY2015,Tan2015,Li2015}. Moreover, we show the band structure along the $k_y$ cut across the Dirac cone near M point in Fig.\ref{FSdirac}(b). It is also in good agreement with the observed bands in experiment\cite{Tan2015,Li2015}, which further confirms the vitality of the obtained nematic order.

\begin{figure}[ht]
\centerline{\includegraphics[height=5cm]{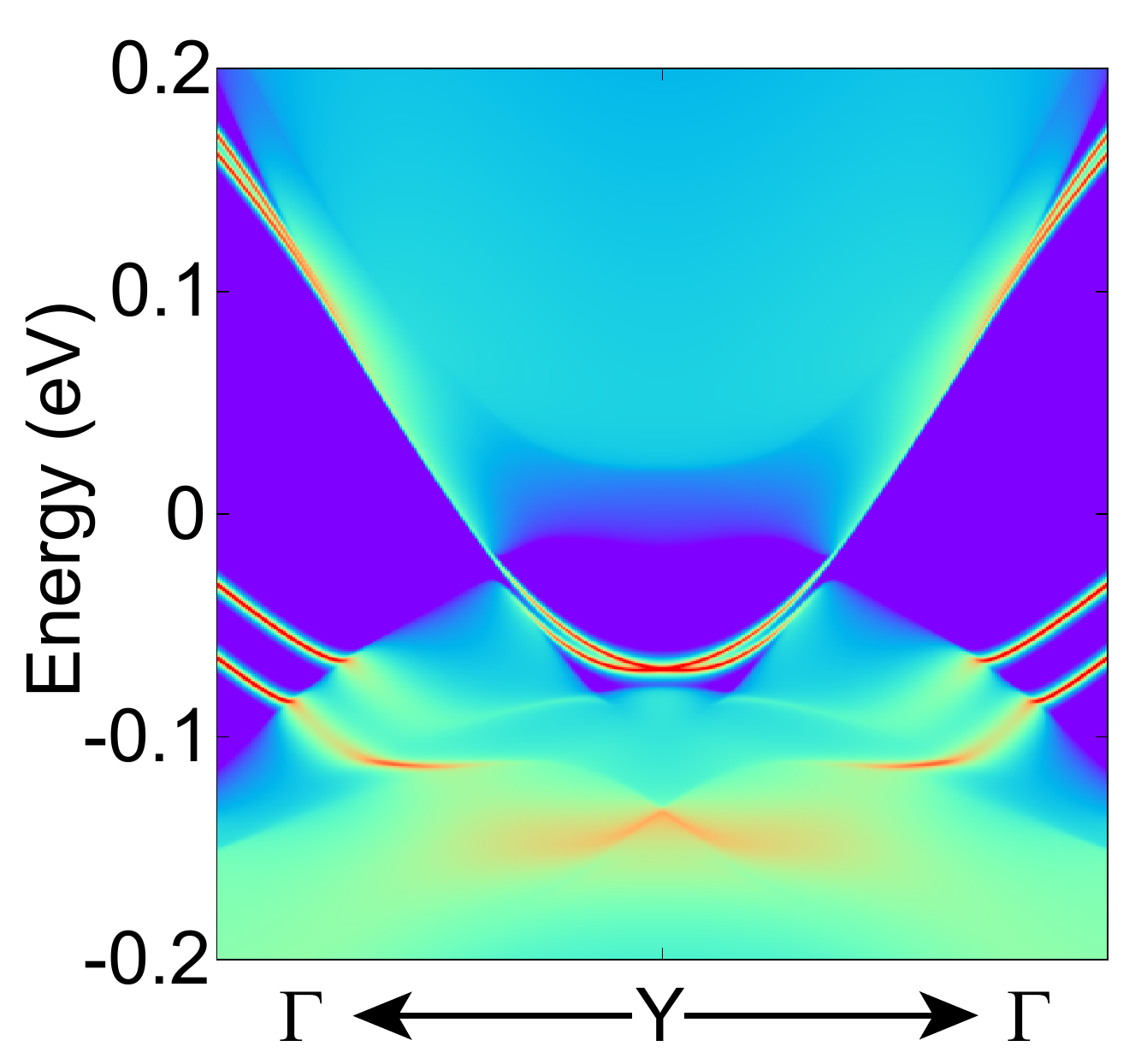}}
\caption{ Energy and momentum dependence of the LDOS for FeSe with nematic order on the [100] edge, where $\chi^A_{1d}=11 $ meV and $\chi^A_{4d}=18$ meV and the spin-orbit coupling strength is 40 meV. The higher LDOS is represented by brighter color. The ingap edge states can be clearly seen around the $Y$ point, which clearly indicates the nontrivial topology.
 \label{edgestate} }
\end{figure}

{\em Nematicity-driven topology}  In normal states, the ``$k$" and ``$k+Q$" bands are degenerate at M point with opposite parities\cite{Wu2014}.  In the nematic states, as the nematic order only breaks the $C_4$ rotational symmetry but not the inversion symmetry,   the parities are still well-defined while the degeneracy is lifted.  Increasing the nematic order, the ``$k+Q$" $d_{xz}$ and $d_{xy}$ bands with opposite parities can move close and eventually cross each other away at $M$ to open a gap. This is a typical band-inversion process required for nontrivial topology in topological insulators. If  spin-orbital coupling is not taken into account,  we expect FeSe  exhibit as a Dirac semimetal when the nematicity pushes the Dirac cones near the FL. With the spin-orbit coupling, the Dirac cones around M point are gapped and the system is topologically nontrivial. Thus, the strength of the nematic order, which can be tuned by applying external pressure, can serve an adjustable  parameter to drive materials to have a topological phase transition.  In Fig.\ref{edgestate}, we show  the existence of  the edge states to confirm the existence of nontrivial topology. The system is  topologically nontrivial if the nematicity-induced splitting is larger than the gap $\Delta^N_M$ between $d_{xz/yz}$ band and $d_{xy}$ band at M point in normal state. In experiment, the splitting between two hole bands (H1, H2) can reach 80 meV\cite{Zhang2015,ZhangY2015} at low temperature, which is much larger than $\Delta^N_M$( less than 40 meV). Thus FeSe in nematic phase is topologically nontrivial.

{\em Discussion}
Our conclusion that the nematic orders are the $d$-wave type of orders in all $t_{2g}$ orbitals have several important consequences.

First,
As we have discussed above,  the $d$-wave nematic orders offer a simple picture  to explain the band structure observed in experiments. The order parameters $\Delta^1_{M}$ of $d_{xz/yz}$ orbitals  and $\Delta^2_{M}$ of $d_{xy}$ orbitals, given in Fig.\ref{bandexp}(a), can be read out from the band structures as the band splitting at M for the two hole-like bands (H1,H2) and for the two electron-like bands (E1,E2) respectively. The appearance of the Dirac cone around M in the nematic state near the Fermi level stems from the arising of the crossing  point of $d_{xz/yz}$ and $d_{xy}$ bands by nematic orders. This explains the experimental observation that the Dirac cone is coexisted with nematicity and disappears when nematicity is suppressed.

 Second, as there are even stronger nematic order in $d_{xy}$ orbitals than in the $d_{xz}$ and $d_{yz}$ orbitals, the nematicity is not simply caused by orbital degeneracy between $d_{xz}$ and $d_{yz}$. Furthermore, the $d$-wave orders are essentially bond orders on the  hopping parameters between $d$-orbitals, which are effectively derived from the coupling to the $p$-orbitals of Se atoms. As the coupling is also responsible for magnetic fluctuations in the materials, the results support that the nematicity and magnetic fluctuations are also strongly entangled. The nematic order may be considered as the key competing order to superconductivity. With increasing electron doping, nematic orders are suppressed to enhance superconductivity. This is consistent with the absence of nematicity  and the appearance strong superconductivity in heavily electron-doped monolayer FeSe on SrTiO$_3$\cite{Tan2015}.

Finally, the existence of the nematicity-induced topological phase transition can help to realize topological physics, such as Majorana modes.  Because of nematicity-induced $C_4$ symmetry breaking, the nematic orders are opposite for two domains  which can be an interesting system to study edge states. These 1D nontrivial edge states are near the FL, as shown in Fig.\ref{edgestate} and can be detected by STM experiments\cite{Yang2012,Drozdov2014}.

In conclusion, we find that the nematic orders in FeSe systems are dominated by the $d$-wave nematic orders in all $t_{2g}$ orbitals. The results can perfectly explain the band structures observed in ARPES experiments in FeSe.  The nematicity can drive a topological phase transition through a band inversion at $M$ point to realize topological physics.

\begin{acknowledgments}
{\em Acknowledgement} We thank useful discussions with D. F. Liu, S. L. He and X. J. Zhou. The work is supported by the Ministry of Science and Technology of China 973 program(Grant No. 2012CV821400), National Science Foundation of China (Grant No. NSFC-1190024,  91536108 and 11104339), and   the Strategic Priority Research Program of  CAS (Grant No. XDB07000000 and  XDB01010000).

\end{acknowledgments}

\end{bibunit}

\begin{bibunit}

\clearpage
\pagebreak
\onecolumngrid
\widetext
\begin{center}
\textbf{\large Supplementary material for ``Nematic orders and nematicity-driven topology in FeSe"}
\end{center}


\setcounter{equation}{0}
\setcounter{figure}{0}
\setcounter{table}{0}
\makeatletter
\renewcommand{\theequation}{S\arabic{equation}}
\renewcommand{\thefigure}{S\arabic{figure}}
\renewcommand{\thetable}{S\arabic{table}}
\renewcommand{\bibnumfmt}[1]{[S#1]}
\renewcommand{\citenumfont}[1]{S#1}

\section{Matrix elements of Hamiltonian and hopping parameters }
\begin{figure}[b]
\centerline{\includegraphics[height=7 cm]{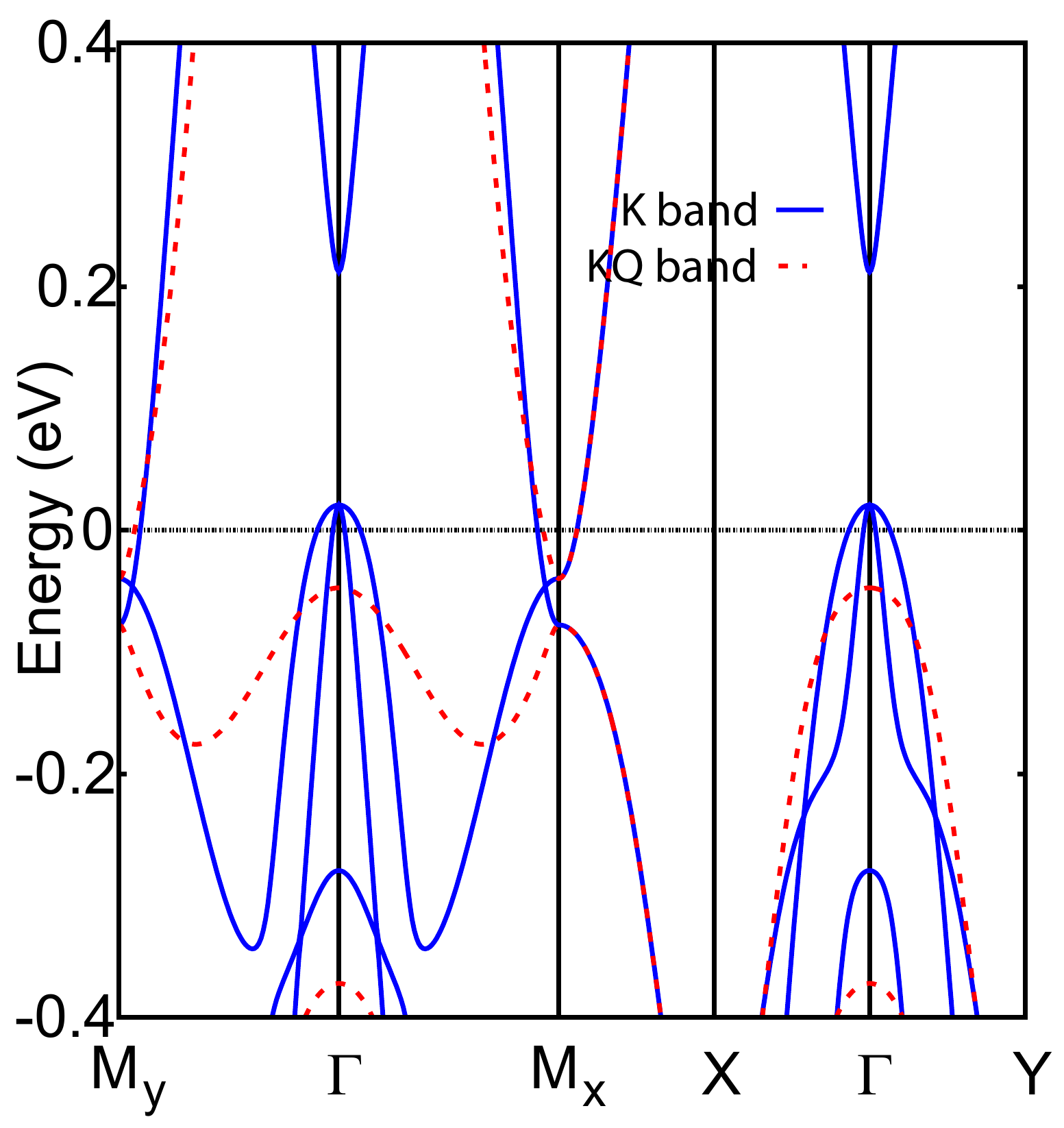}}
\caption{(color online). The final band structure of FeSe. The blue solid lines represent the "k" band and the red dahsed lines represent the "k+Q" band.  \label{bandkkQ} }
\end{figure}

\begin{figure}[tb]
\centerline{\includegraphics[height=6 cm]{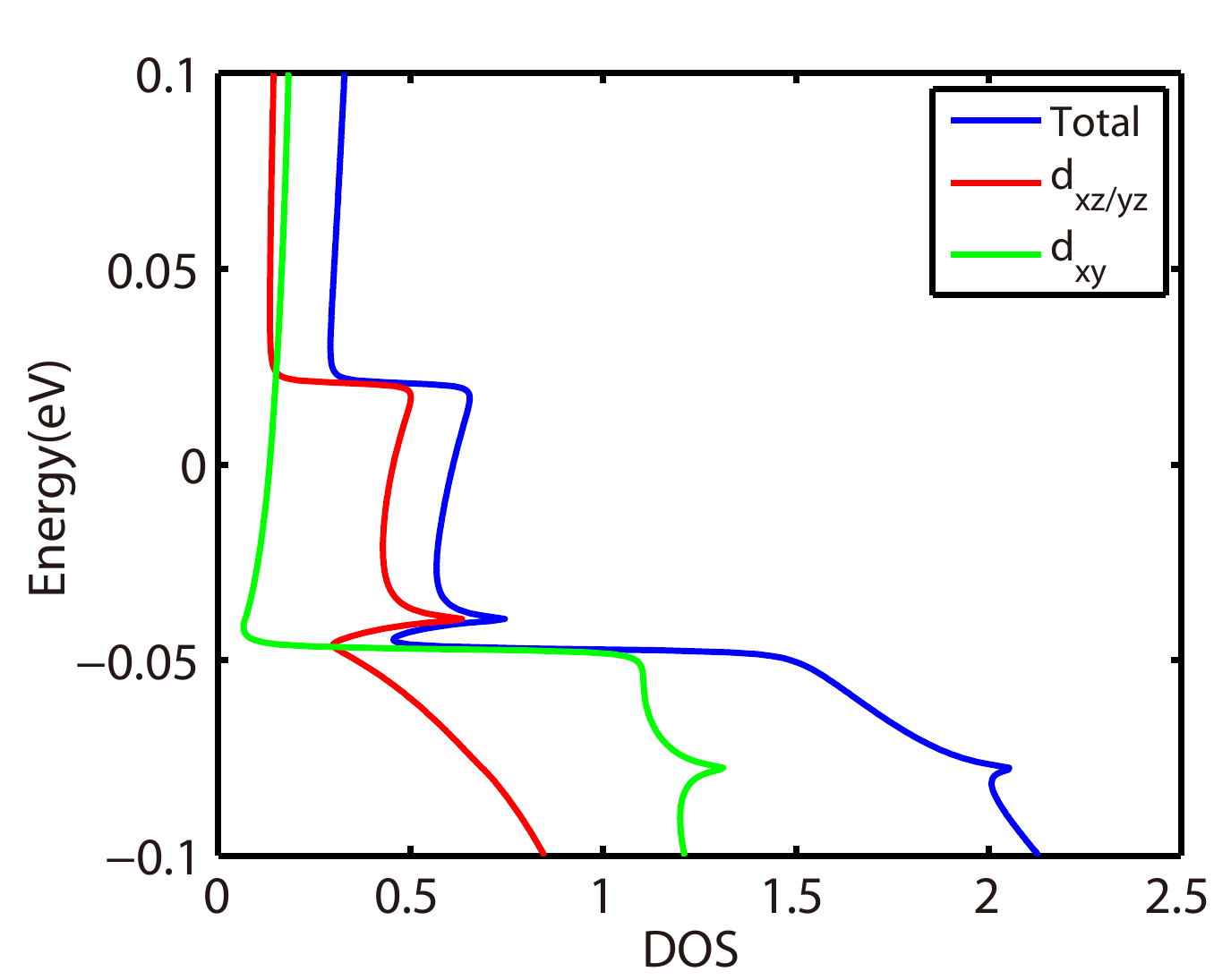}}
\caption{(color online). Density of states of the band structure.   \label{dos} }
\end{figure}

The final band structure with renormalization and additional shifts to hopping parameters is given in Fig.\ref{bandkkQ}, where the blue solid lines denote "k" band and the red dashed lines denote the "k+Q" band. The matrix elements of $A(\mathbf{k})$ are in the following:
\begin{eqnarray*}
e_{11/22}(\mathbf{k}) & = &
  \epsilon_{11/22}+2t_{x/y}^{11}cosk_{x}+2t_{y/x}^{11}cosk_{y}+4t_{xy}^{11}cosk_{x}cosk_{y}+2t_{xx/yy}^{11}cos2k_{x}+2t_{yy/xx}^{11}cos2k_{y}\\
 &  &
  +4t_{xyy/xxy}^{11}cosk_{x}cos2k_{y}+4t_{xxy/xyy}^{11}cos2k_{x}cosk_{y}+4t_{xxyy}^{11}cos2k_{x}cos2k_{y},\\
e_{33/44/55}(\mathbf{k}) & = &
  \epsilon_{33/44/55}+2t_{x}^{3/4/5}(cosk_{x}+cosk_{y})+4t_{xy}^{33/44/55}cosk_{x}cosk_{y}+2t_{xx}^{33/44/55}(cos2k_{x}+cos2k_{y})\\
 &  & +4t_{xxy}^{33/44/55}(cosk_{x}cos2k_{y}+cos2k_{x}cosk_{y})+4t_{xxyy}^{33/44/55}cos2k_{x}cos2k_{y},\\
e_{12}(\mathbf{k}) & = &
  -4t_{xy}^{12}sink_{x}sink_{y}-4t_{xxy}^{12}(sink_{x}sin2k_{y}-sin2k_{x}sink_{y})-4t_{xxyy}^{44}sin2k_{x}sin2k_{y},\\
e_{13/23}(\mathbf{k}) & = &
  \pm2it_{y}^{13}sink_{y/x}\pm4it_{xy}^{13}cosk_{x/y}sink_{y/x}\pm2it_{yy}^{13}sin2k_{y/x}\\
 &  & \pm4it_{xxy}^{13}cos2k_{x/y}sink_{y/x}\pm4it_{xyy}^{13}cosk_{x/y}sin2k_{y/x},\\
e_{14/24}(\mathbf{k}) & = &
  2it_{x}^{14}sink_{x/y}+4it_{xy}^{14}sink_{x/y}cosk_{y/x}+2it_{xx}^{14}sin2k_{x/y}\\
 &  &+4it_{xxy}^{14}sin2k_{x/y}cosk_{y/x}+4it_{xxyy}^{14}sin2k_{x/y}cos2k_{y/x},\\
e_{15/25}(\mathbf{k}) & = &
  2it_{y}^{15}sink_{y/x}+4it_{xy}^{15}cosk_{x/y}sink_{y/x}+2it_{yy}^{15}sin2k_{y/x}\\
 &  & +4it_{xxy}^{15}cos2k_{x/y}sink_{y/x}+4it_{xyy}^{15}cosk_{x/y}sin2k_{y/x}+4it_{xxyy}^{15}cos2k_{x/y}sin2k_{y/x},\\
e_{34}(\mathbf{k}) & = & -4t_{xyy}^{34}(sink_{x}sin2k_{y}-sin2k_{x}sink_{y}),\\
e_{35}(\mathbf{k}) & = &
   2t_{x}^{35}(cosk_{x}-cosk_{y})+2t_{xx}^{35}(cos2k_{x}-cosk_{2y})+4t_{xxy}^{35}(cosk_{x}cos2k_{y}-cos2k_{x}cosk_{y}),\\
e_{45}(\mathbf{k}) & = & -4t_{xy}^{45}sink_{x}sink_{y}-4t_{xxyy}^{45}sin2k_{x}sink_{y}.
\end{eqnarray*}
The corresponding hopping parameters are given in Table \ref{hopping}. The density of states is shown in Fig.\ref{dos}, where the Fermi level is close to the van Hove singularity contributed by $d_{xz/yz}$ and $d_{xy}$ orbitals.

\begin{table}[bt]
\caption{\label{hopping}Final Hopping parameters for monolayer FeSe.  The $x$ direction is along the Fe-Fe bond. The onsite energies of $d$ oribtals are (all in eV) : $\epsilon_1$=0.1754, $\epsilon_3$=-0.3576, $\epsilon_4$=0.0904, $\epsilon_5$=-0.2776.}
\begin{ruledtabular}
\begin{tabular}{ccccccccc}
  $t^{mn}_i$ & $i$=$x$ & $i$=$y$ & $i$=$xy$ & $i$=$xx$ & $i$=$yy$ & $i$=$xxy$ & $i$=$xyy$ & $i$=$xxyy$ \\
 \colrule
  $mn$=11 &  -0.1344  & -0.4009 & 0.227  & 0.002 & -0.036 & -0.019 & 0.014 & 0.024      \\
  $mn$=33 & 0.4584  &        & -0.070 & -0.013 &        &        &       &  0.012      \\
  $mn$=44 &0.0704  &        & 0.0200 & 0.002 &        &  -0.019      &       &  -0.024       \\
  $mn$=55 &   &        & 0.013 &  -0.014   &      & 0.006 &   &  -0.011      \\
  $mn$=12 &        &        & 0.103  &        &        &  -0.011      &       & 0.032       \\
  $mn$=13 &        & 0.473  &  -0.089      &        &  0.011      &  -0.018      &   0.006    &  \\
  $mn$=14 & 0.2736 &        & 0.053  &    -0.001    &        &   0.006     &       &  -0.009     \\
  $mn$=15 &        & 0.2   & -0.13  &        &    0.009    &  -0.009 &  -0.011     & -0.012       \\
  $mn$=34 &        &        &        &        &   &  0.012      &       &       \\
  $mn$=35 & -0.401 &        &        &  -0.023      &        &  -0.006      &       &       \\
  $mn$=45 &        &        & -0.113 &        &        &        &       &    0.011    \\
\end{tabular}
\end{ruledtabular}

\end{table}

\section{Nearest Neighbor interaction}

According to eigenvalues of the glide-plane operation, the $d$ orbitals can be classified into two groups: $g_1$ ($d_{xz}$, $d_{yz}$) and $g_2$ ($d_{xy}$, $d_{x^2-y^2}$, $d_{z^2}$). In the tight binding model, the intragroup orbitals couple with each other with the same momentum and the intergroup oribtals couple through the term $c^{\dag}_{kg_1}c_{k+Qg_2}$. With respect to the glide-plane symmetry, consider the bond order by assuming,
 \begin{eqnarray}\langle c^{\dag}_{i\alpha\sigma}c_{i+x/y\beta\sigma'}\rangle=
\begin{cases}
n^{x/y}_{\alpha\beta}\delta_{\sigma\sigma'}, & \alpha \in g_\alpha, \beta \in g_\beta, g_\alpha = g_\beta,  \cr
-e^{i\mathbf{i}\cdot \mathbf{Q}}n^{\alpha\beta}_{x/y}\delta_{\sigma\sigma'} &\alpha \in g_\alpha, \beta \in g_\beta, g_\alpha \neq g_\beta.  \cr
\end{cases}
\end{eqnarray}
In the mean-field level, the total interaction can be rewritten as,
\begin{eqnarray}
H_V&=&-V\sum^5_{\alpha=1}\sum_{k\sigma}(2n^{\alpha}_xcosk_x+2n^{\alpha}_{y}cosk_y)c^{\dag}_{k\alpha\sigma}c_{k\alpha\sigma}+VN\sum^5_{\alpha=1}\sum_{\sigma}(|n^{\alpha}_{x\sigma}|^2+|n^{\alpha}_{y\sigma}|^2)\nonumber\\
&&-V\sum_{ k\sigma }[2i n^{13}_ysink_yc^{\dag}_{k1\sigma}c_{k+Q3\sigma}-2i n^{13}_{y}sink_y c^{\dag}_{k+Q3\sigma}c_{k1\sigma}]+\sum_{\sigma}2VN|n^{13}_{y\sigma}|^2\nonumber\\
&&-V\sum_{ k\sigma}[2i n^{23}_xsink_xc^{\dag}_{k2\sigma}c_{k+Q3\sigma}-2i n^{23}_{x}sink_x c^{\dag}_{k+Q3\sigma}c_{k2\sigma}]+\sum_{\sigma}2VN|n^{23}_{x\sigma}|^2\nonumber\\
&&-V\sum_{ k\sigma}[2i n^{14}_xsink_xc^{\dag}_{k1\sigma}c_{k+Q4\sigma}-2i n^{14}_{x}sink_x c^{\dag}_{k+Q4\sigma}c_{k1\sigma}]+\sum_{\sigma}2VN|n^{14}_{x\sigma}|^2\nonumber\\
&&-V\sum_{ k\sigma}[2i n^{24}_ysink_yc^{\dag}_{k2\sigma}c_{k+Q4\sigma}-2i n^{24}_{y}sink_y c^{\dag}_{k+Q4\sigma}c_{k2\sigma}]+\sum_{\sigma}2VN|n^{24}_{y\sigma}|^2\nonumber\\
&&-V\sum_{ k\sigma }[2i n^{15}_ysink_yc^{\dag}_{k1\sigma}c_{k+Q5\sigma}-2i n^{15}_{y}sink_y c^{\dag}_{k+Q5\sigma}c_{k1\sigma}]+\sum_{\sigma}2VN|n^{15}_{y\sigma}|^2\nonumber\\
&&-V\sum_{ k\sigma }[2i n^{25}_xsink_xc^{\dag}_{k2\sigma}c_{k+Q5\sigma}-2i n^{25}_{x}sink_x c^{\dag}_{k+Q5\sigma}c_{k2\sigma}]+\sum_{\sigma}2VN|n^{25}_{x\sigma}|^2\nonumber\\
&&-V\sum_{k\sigma}(2n^{35}_xcosk_x+2n^{35}_ycosk_y)(c^{\dag}_{k3\sigma}c_{k5\sigma}+c^{\dag}_{k5\sigma}c_{k3\sigma})+\sum_{\sigma}2VN(|n^{35}_{x\sigma}|^2+|n^{35}_{y\sigma}|^2)
\end{eqnarray}
The above Hamiltonian can be further written as,
\begin{eqnarray}
H_V&=& E_0+\sum_{k\sigma}2\chi^S_{1s}(cosk_x+cosk_y)(c^{\dag}_{k1\sigma}c_{k1\sigma}+c^{\dag}_{k2\sigma}c_{k2\sigma})+\sum_{k\sigma}2\chi^S_{1d}(cosk_x-cosk_y)(c^{\dag}_{k1\sigma}c_{k1\sigma}-c^{\dag}_{k2\sigma}c_{k2\sigma})\nonumber\\
&&+\sum_{k\sigma}2\chi^A_{1s}(cosk_x+cosk_y)(c^{\dag}_{k1\sigma}c_{k1\sigma}-c^{\dag}_{k2\sigma}c_{k2\sigma})+\sum_{k\sigma}2\chi^A_{1d}(cosk_x-cosk_y)(c^{\dag}_{k1\sigma}c_{k1\sigma}+c^{\dag}_{k2\sigma}c_{k2\sigma})\nonumber\\
&&+\sum^5_{\alpha=3}\sum_{k\sigma}[2\chi^S_{\alpha s}(cosk_x+cosk_y)+2\chi^A_{\alpha d}(cosk_x-cosk_y)]c^{\dag}_{k\alpha\sigma}c_{k\alpha\sigma}\nonumber\\
&&+\sum_{ k\sigma }[2i \chi^S_{13p}sink_yc^{\dag}_{k1\sigma}c_{k+Q3\sigma}-2i \chi^S_{13p}sink_xc^{\dag}_{k2\sigma}c_{k+Q3\sigma}+h.c.]\nonumber\\
&&+\sum_{ k\sigma }[2i \chi^A_{13p}sink_yc^{\dag}_{k1\sigma}c_{k+Q3\sigma}+2i \chi^A_{13p}sink_xc^{\dag}_{k2\sigma}c_{k+Q3\sigma}+h.c.] \nonumber\\
&&+\sum_{ k\sigma }[2i \chi^S_{14p}sink_xc^{\dag}_{k1\sigma}c_{k+Q4\sigma}+2i \chi^S_{14p}sink_yc^{\dag}_{k2\sigma}c_{k+Q4\sigma}+h.c.]\nonumber\\
&&+\sum_{ k\sigma }[2i \chi^A_{14p}sink_xc^{\dag}_{k1\sigma}c_{k+Q4\sigma}-2i \chi^A_{14p}sink_yc^{\dag}_{k2\sigma}c_{k+Q4\sigma}+h.c.]\nonumber\\
&&+\sum_{ k\sigma }[2i \chi^S_{15p}sink_yc^{\dag}_{k1\sigma}c_{k+Q5\sigma}+2i \chi^S_{15p}sink_xc^{\dag}_{k2\sigma}c_{k+Q5\sigma}+h.c.]\nonumber\\
&&+\sum_{ k\sigma }[2i \chi^A_{15p}sink_yc^{\dag}_{k1\sigma}c_{k+Q5\sigma}-2i \chi^A_{15p}sink_xc^{\dag}_{k2\sigma}c_{k+Q5\sigma}+h.c.]\nonumber\\
&&+\sum_{k\sigma}[2\chi^S_{35d}(cosk_x-cosk_y)+2\chi^A_{35s}(cosk_x+cosk_y)](c^{\dag}_{k3\sigma}c_{k5\sigma}+c^{\dag}_{k5\sigma}c_{k3\sigma}),
\end{eqnarray}
where $\chi^S$ represents the symmetry preserving order parameter and $\chi^A$ the symmetry breaking order parameter. The order parameters are given by,
\begin{eqnarray}
&&\chi^S_{1s}=-\frac{V}{4}(n^1_x+n^1_y+n^2_x+n^2_y),\chi^A_{1s}=-\frac{V}{4}(n^1_x+n^1_y-n^2_x-n^2_y)\nonumber\\
&&\chi^S_{1d}=-\frac{V}{4}(n^1_x-n^1_y+n^2_x-n^2_y),\chi^A_{1d}=-\frac{V}{4}(n^1_x-n^1_y-n^2_x+n^2_y) \nonumber\\
&&\chi^S_{\alpha s}=-\frac{V}{2}(n^\alpha_x+n^\alpha_y),\chi^A_{\alpha d}=-\frac{V}{2}(n^\alpha_x-n^\alpha_y) (\alpha=3,4,5),\nonumber\\
&& \chi^S_{13p}=-\frac{V}{2}(n^{13}_y-n^{23}_x), \chi^A_{13p}=-\frac{V}{2}(n^{13}_y+n^{23}_x),\nonumber\\
&& \chi^S_{14p}=-\frac{V}{2}(n^{14}_x+n^{24}_y), \chi^A_{14p}=-\frac{V}{2}(n^{14}_x-n^{24}_y),\nonumber\\
&& \chi^S_{15p}=-\frac{V}{2}(n^{15}_y+n^{25}_x), \chi^A_{15p}=-\frac{V}{2}(n^{15}_y-n^{25}_x),\nonumber\\
&&\chi^A_{35s}=-\frac{V}{2}(n^{35}_x+n^{35}_y), \chi^S_{35d}=-\frac{V}{2}(n^{35}_x-n^{35}_y).
\end{eqnarray}
The intra-orbital order parameter is,
\begin{eqnarray}
n^{\alpha}_{x/y}=\frac{1}{N}\sum_i \langle c^{\dag}_{i+x/y\alpha}c_{i\alpha} \rangle=\frac{1}{N}\sum_k cosk_{x/y} \langle c^{\dag}_{k\alpha}c_{k\alpha} \rangle.
\end{eqnarray}
The inter-orbital order parameters are,
\begin{eqnarray}
n^{14}_{x}&=&\frac{1}{N}\sum_k isink_x \langle c^{\dag}_{k1\uparrow}c_{k+Q4\uparrow}\rangle \\
n^{24}_{y}&=&\frac{1}{N}\sum_k isink_y \langle c^{\dag}_{k2\uparrow}c_{k+Q4\uparrow}\rangle\\
n^{13}_{y}&=&\frac{1}{N}\sum_k isink_y \langle c^{\dag}_{k1\uparrow}c_{k+Q3\uparrow}\rangle\\
n^{23}_{x}&=&\frac{1}{N}\sum_k isink_x \langle c^{\dag}_{k2\uparrow}c_{k+Q3\uparrow}\rangle \\
n^{15}_{y}&=&\frac{1}{N}\sum_k isink_y \langle c^{\dag}_{k1\uparrow}c_{k+Q5\uparrow}\rangle\\
n^{25}_{x}&=&\frac{1}{N}\sum_k isink_x \langle c^{\dag}_{k2\uparrow}c_{k+Q5\uparrow}\rangle
\end{eqnarray}
The Energy constant is,
\begin{eqnarray}
E_0&=&\frac{4N}{V}\sum_{\sigma}(|\chi^S_{1s}|^2+|\chi^S_{1d}|^2+|\chi^A_{1s}|^2+|\chi^A_{1d}|^2)+\frac{2N}{V}\sum_{\sigma}(|\chi^S_{3s}|^2+|\chi^A_{3d}|^2)\nonumber\\
&&+\frac{2N}{V}\sum_{\sigma}(|\chi^S_{4s}|^2+|\chi^A_{4d}|^2)+\frac{2N}{V}\sum_{\sigma}(|\chi^S_{5s}|^2+|\chi^A_{5d}|^2)+\frac{4N}{V}\sum_{\sigma}(|\chi^S_{13p}|^2+|\chi^A_{13p}|^2)\nonumber\\
&&+\frac{4N}{V}\sum_{\sigma}(|\chi^S_{14p}|^2+|\chi^A_{14p}|^2)+\frac{4N}{V}\sum_{\sigma}(|\chi^S_{15p}|^2+|\chi^A_{15p}|^2)+\frac{4N}{V}\sum_{\sigma}(|\chi^S_{35d}|^2+|\chi^A_{35s}|^2)
\end{eqnarray}

\end{bibunit}

\end{document}